\documentclass[a4paper,fleqn,usenatbib,useAMS]{mnras}

\usepackage{mathptmx}

\usepackage[T1]{fontenc}
\usepackage{ae,aecompl}

\usepackage{graphicx}	
\usepackage{amsmath}	
\usepackage{amssymb}	

\bibpunct{(}{)}{;}{a}{}{,}
\newcommand{\logg}{\log\,g}
\newcommand{\teff}{T_{\rm eff}}
\newcommand{\kms}{km\,s$^{-1}$}
\newcommand{\masyr}{mas\,yr$^{-1}$}

\title[Two magnetic pulsating stars]{
  A magnetic study of the $\delta$\,Scuti variable HD\,21190 and the close
  solar-type background star CPD\,$-$83$^{\circ}$\,64B}

\author[S. P. J\"arvinen et al.]{
S.~P.~J\"arvinen$^{1}$\thanks{E-mail: sjarvinen@aip.de},
S.~Hubrig$^{1}$,
R.~-D.~Scholz$^{1}$,
E.~Niemczura$^{2}$,
I.~Ilyin$^{1}$,
\and
M.~Sch\"oller$^{3}$
\\
$^{1}$Leibniz-Institut f\"ur Astrophysik Potsdam (AIP),
An der Sternwarte~16, 14482~Potsdam, Germany\\
$^{2}$Instytut Astronomiczny, Uniwersytet Wroc{\l}awski,
Kopernika 11, 51-622 Wroc{\l}aw, Poland\\
$^{3}$European Southern Observatory,
Karl-Schwarzschild-Str.~2, 85748~Garching, Germany\\
}

\date{Accepted 2018. Received 2018; in original form 2018}

\pubyear{2018}

\begin{document}
\label{firstpage}
\pagerange{\pageref{firstpage}--\pageref{lastpage}}
\maketitle

\begin{abstract}
  HD\,21190 is a known $\delta$\,Scuti star showing Ap star characteristics and
  a variability period of 3.6\,h discovered by the \emph{Hipparcos} mission.
  Using {\it Gaia} DR1 data for an astrometric analysis, it was recently
  suggested that HD\,21190 forms a physical binary system with the companion
  CPD\,$-$83$^{\circ}$\,64B. An atmospheric chemical analysis based on HARPS
  observations revealed the presence of overabundances of heavy and rare-earth
  elements, which are typically observed in chemically peculiar stars with
  large-scale organized magnetic fields. Previous observations of HD\,21190
  indicated a magnetic field strength of a few hundred Gauss. The presence of
  a magnetic field in CPD\,$-$83$^{\circ}$\,64B remained unexplored. In this
  work, we reanalyse this system using {\it Gaia} DR2 data and
  present our search for the magnetic field in both stars based on multi-epoch
  HARPS\-pol high-resolution and FORS\,2 low-resolution spectropolarimetric
  observations. The {\it Gaia} DR2 results clearly indicate that the two stars
  are not physically associated.
  A magnetic field detection at a significance level of more than 6$\sigma$
  ($\left< B_{\rm z}\right>_{\rm all}=230\pm38$\,G) was achieved for the
  $\delta$\,Scuti variable HD\,21190 in FORS\,2 observations using the entire
  spectrum for the measurements. The magnetic field appears to be stronger in
  CPD\,$-$83\degr\,64B. The highest value for the longitudinal magnetic field
  in CPD\,$-$83$^{\circ}$\,64B, $\left< B_{\rm z}\right>_{\rm all}=509\pm104$\,G,
  is measured at a significance level of 4.9$\sigma$. Furthermore, the
  high-resolution HARPS\-pol observations of this component indicate the
  presence of pulsational variability on a time scale of tens of minutes.
 \end{abstract}

\begin{keywords}
  stars: individual: HD\,21190, CPD\,$-$83\degr\,64B --
  stars: magnetic field --
  stars: oscillations
\end{keywords}



\section{Introduction}
\label{sec:intro}

HD\,21190 is a known $\delta$~Scuti variable star showing Ap star
characteristics. Using time-series photometry obtained by the \emph{Hipparcos}
mission,
\citet{Koen2001}
discovered a variability period of 3.6\,h. According to these authors, the
spectral classification of HD\,21190 is F2\,III SrEuSi, making it the most
evolved Ap star known.
\citet{Gonzales2008}
used the
Ultraviolet and Visual Echelle Spectrograph
\citep[UVES;][]{uves}
time-series of HD\,21190 to search for pulsational line profile variations and
were able to show that this star presents the best example of a $\delta$~Scuti
star with moving bumps in its line profiles, which is characteristic for
high-degree pulsations. The weak longitudinal magnetic field of HD\,21190,
$\left< B_{\rm z}\right>_{\rm hyd}=47\pm13$\,G, was detected by
\citet{Kurtz2008}
from FOcal Reducer low dispersion Spectrograph
\citep[FORS\,1;][]{1998Msngr}
observations in spectropolarimetric mode. Re-observation of this target,
$\left< B_{\rm z}\right>_{\rm hyd}=$-$237\pm75$\,G and
$\left< B_{\rm z}\right>_{\rm all}=$-$254\pm60$\,G using FORS\,2
\citep{2016HubSch}
with the same setup as
\citet{Kurtz2008}
confirmed the magnetic field detection.

A recent astrometric analysis
\citep{niemczura}
using {\it Gaia} DR1 data
\citep{gaia16}
showed that the $\delta$~Scuti variable HD\,21190 forms a physical binary
system with CPD\,$-$83\degr\,64B at an angular distance of about 19\arcsec. An
atmospheric chemical analysis based on observations with the High Accuracy
Radial velocity Planet Searcher polarimeter
\citep[HARPS\-pol;][]{snik2008}
yielded the abundances of 24 chemical elements in the atmosphere of HD\,21190
and 28 chemical elements in CPD\,$-$83\degr\,64B.
\citet{niemczura}
obtained the atmospheric parameters $\teff=6900\pm100$\,K and
$\logg=3.60\pm0.20$ for HD\,21190 and $\teff=5850\pm50$\,K and
$\logg=4.3\pm0.1$ for CPD\,$-$83\degr\,64B. The abundance analysis of
HD\,21190 showed that all the studied heavy and rare-earth elements appear
overabundant, confirming the peculiar nature of this star. The abundance
pattern for CPD\,$-$83\degr\,64B appears mostly solar with modest enhancements
of heavy and rare-earth elements.
  
With the most recent release of the {\it Gaia} DR2 data
\citep{gaia18},
it became possible to re-investigate the physical connection between the two
stars. In the following, we present a brief analysis of these results.
Further, while a few spectropolarimetric observations were obtained in the
past for HD\,21190, the presence of a magnetic field in CPD\,$-$83\degr\,64B
remained unexplored. We present our recently acquired multi-epoch high- and
low-resolution spectropolarimetric observations for both objects.
 

\section{{\it Gaia} DR2 results}
\label{Sect_gaiadr2}

\begin{table}
\centering
\caption{{\it Gaia} DR2 astrometry, radial velocities, and derived physical
  parameters of HD\,21190 and CPD\,$-$83$^{\circ}$\,64B.}
\label{tab:table_gaiadr2}
\begin{tabular}{lc|cc}
\hline
Object           &           & HD\,21190          & CPD\,$-$83$^{\circ}$\,64B \\
\hline
RA$^{a}$         & [degrees]  & 47.473840         & 47.432566 \\
DE$^{a}$         & [degrees] & $-$83.531459       & $-$83.528743 \\
$Plx$            & [mas]    & 4.9811$\pm$0.0492  & 4.0170$\pm$0.0542 \\
distance$^{b}$   & [pc]      & 199.6$^{+2.0}_{-2.0}$ & 247.2$^{+3.4}_{-3.3}$ \\
$pm$RA           & [\masyr] & $-$6.152$\pm$0.127  & $-$0.210$\pm$0.143 \\
$pm$DE           & [\masyr] & $+$21.742$\pm$0.093 & $+$24.116$\pm$0.098 \\
$RV$             & [\kms]   & -                   & $+$26.06$\pm$3.43 \\
$T_{\rm eff}$$^{c}$ & [K]      & 6798$^{+97}_{-62}$    & 5800$^{+122}_{-135}$ \\
radius           & $R_{\sun}$ & 3.85$^{+0.07}_{-0.11}$ & 1.44$^{+0.07}_{-0.06}$ \\
\hline
\end{tabular}
\begin{minipage}{\columnwidth}
{\bf Notes:}\\
$^{a}$ {\it Gaia} DR2 coordinates are for (J2000, epoch 2015.5) and were
rounded to 0.000001 degrees;
$^{b}$ estimated distances with their lower and upper bounds of their
confidence intervals according to \citet{bailer-jones18} were rounded to
0.1\,pc;
$^{c}$ {\it Gaia} DR2 effective temperatures and their uncertainties were
rounded to 1\,K.
\end{minipage}
\end{table}

Based on {\it Gaia} DR1 data,
\citet{niemczura}
considered HD\,21190 and CPD\,$-$83$^{\circ}$\,64B to be a wide binary system.
However, the new {\it Gaia} DR2
\citep{gaia18}
data clearly rule out a physical connection between these two objects and show
that CPD\,$-$83$^{\circ}$\,64B is an unrelated background star appearing close
to HD\,21190. The {\it Gaia} DR2 data of HD\,21190 and
CPD\,$-$83$^{\circ}$\,64B, including parallaxes (and distance estimates by
\citet{bailer-jones18}), 
proper motions, the radial velocity of CPD\,$-$83$^{\circ}$\,64B, and derived
physical parameters, are listed in Table~\ref{tab:table_gaiadr2}. These data
are discussed in the following subsections.

\subsection{Astrometry}
\label{SubSect_astm}

The physical association of HD\,21190 and CPD\,$-$83$^{\circ}$\,64B previously
asserted by
\citet{niemczura}
was mainly based on the common proper motion derived by these authors using
the same three epochs of modern non-photographic positional measurements for
both sources, including {\it Gaia} DR1
\citep{gaia16}
positions. The corresponding proper motions agreed within their error bars
(about $\pm$2\,\masyr). 
\citet{niemczura}
also considered the proper motions published in two different catalogues
involving {\it Gaia} DR1 data, namely the 5th United States Naval Observatory
CCD Astrograph Catalog (UCAC5) of
\citet{zacharias17} 
and the Tycho-{\it Gaia} Astrometric Solution (TGAS) of
\citet{lindegren16}.
The difference between the TGAS proper motion of HD\,21190 and the UCAC5
proper motion of CPD\,$-$83$^{\circ}$\,64B was found to be even smaller than
1\,\masyr.

Whereas the {\it Gaia} DR2 proper motion of HD\,21190 deviates by less than
about 0.7\,\masyr{} from the above mentioned previous proper motion
measurements, large discrepancies of 6--7\,\masyr\ in RA and 2--5\,\masyr{} in
DE are found for CPD\,$-$83$^{\circ}$\,64B (see our
Table~\ref{tab:table_gaiadr2} and Table~1 in
\citet{niemczura}).
With the much higher precision of the {\it Gaia} DR2 proper motions (errors of
the order of 0.1\,\masyr{} for both stars), the two stars can not be
considered a common proper motion pair any longer.

The difference in their {\it Gaia} DR2 parallaxes is about 20 times larger
than the formal errors (see Table~\ref{tab:table_gaiadr2}). By simply
inverting the parallaxes, we derive distances of 200.8\,pc for HD\,21190 and
248.9\,pc for CPD\,$-$83$^{\circ}$\,64B, very similar to the distance estimates
of
\citet{bailer-jones18}
given in Table~\ref{tab:table_gaiadr2}. Therefore, CPD\,$-$83$^{\circ}$\,64B
lies about 50\,pc further away than HD\,21190. The new distances of both
objects are much larger than the TGAS-based distance of HD\,21190 (165\,pc)
that was used by
\citet{niemczura}
as their assumed common distance.

We have also checked how reliable the astrometric data of both stars are with
respect to various {\it Gaia} DR2 quality flags as discussed by
\citet{lindegren18}.
Although the two stars lie beyond the 100\,pc horizon used by
\citet{lindegren18}
for defining a high-quality sample (their ''Selection C'') of 242,582 objects
within 100\,pc, they otherwise fulfil all their selection criteria applied to 
astrometric and photometric quality flags.

\subsection{Radial velocity}
\label{SubSect_RV}

A radial velocity measurement is reported in {\it Gaia} DR2 for only one of
the two stars, CPD\,$-$83$^{\circ}$\,64B (see Table~\ref{tab:table_gaiadr2}).
This radial velocity is only slightly larger than the value of
$+$19$\pm$0.5\,\kms{} measured by
\citet{niemczura}.
On the other hand, a much larger radial velocity of $+$34.0$\pm$1.3\,\kms{}
was reported by the Radial Velocity Experiment (RAVE) DR5
\citep{kunder17}.
As we found no hint on possible problems with the RAVE measurements of this
star, it is possible that CPD\,$-$83$^{\circ}$\,64B is a close binary
unresolved in {\it Gaia} DR2.

\subsection{Physical parameters}
\label{SubSect_physpar}

Astrophysical parameters, including stellar effective temperatures, radii and
luminosities, were derived for 77 million sources
\citep{andrae18}
by using the {\it Gaia} DR2 photometry and parallaxes. These parameters are
part of {\it Gaia} DR2. In Table~\ref{tab:table_gaiadr2} we show the
{\it Gaia} DR2 effective temperatures and radii of HD\,21190 and
CPD\,$-$83$^{\circ}$\,64B. The effective temperatures are in good agreement
with the values estimated by
\citet{niemczura}
in their analysis of the chemical abundances using high-resolution HARPS\-pol
spectra.


\section{Observations}
\label{sec:obs}

\begin{table}
\centering
\caption{
Logbook of all individual subexposures of the spectropolarimetric
observations of HD\,21190 (left) and CPD\,$-$83\degr\,64B (right) with
HARPS\-pol. The columns give the heliocentric Julian date of mid-exposure, the
exposure time, and the achieved S/N in the Stokes $I$ spectra around
6450\,\AA{}.
}
\label{tab:log}
\begin{tabular}{ccrccr}
\hline
\multicolumn{3}{c}{HD\,21190}
& 
\multicolumn{3}{c}{CPD\,$-$83\degr\,64B} \\
\multicolumn{1}{c}{HJD} &
\multicolumn{1}{c}{Exp.} &
\multicolumn{1}{c}{S/N} & 
\multicolumn{1}{c}{HJD} &
\multicolumn{1}{c}{Exp.} &
\multicolumn{1}{c}{S/N} \\
\multicolumn{1}{c}{2\,450\,000+} &
\multicolumn{1}{c}{time} & & 
\multicolumn{1}{c}{2\,450\,000+} &
\multicolumn{1}{c}{time} & \\
 &
\multicolumn{1}{c}{(s)} & & 
 &
 \multicolumn{1}{c}{(s)} & \\
\hline
57554.9222 & 1500 &  83 & 57555.8892 & 1800 &  50 \\
57554.9400 & 1500 &  79 & 57555.9104 & 1800 &  41 \\
57557.9197 & 1900 & 193 & 57555.9316 & 1800 &  48 \\
57557.9409 & 1700 & 209 & 57555.9488 & 1100 &  25 \\
57558.9206 & 1450 & 116 &            &      &  \\
57558.9372 & 1360 & 124 &            &      &  \\
\hline
\end{tabular}
\end{table}

The high-resolution spectropolarimetric observations of HD\,21190 and
CPD\,$-$83\degr\,64B were obtained with HARPS\-pol attached to ESO's 3.6\,m
telescope (La Silla, Chile). HD\,21190 was observed on 2016 June 15, 18, and
19, while CPD\,$-$83\degr\,64B was observed on 2016 June 16. Each observation
consisted of subexposures with exposure times varying between 23 and 32\,min
for HD\,21190 and between 18 and 30\,min for CPD\,$-$83\degr\,64B. The
quarter-wave retarder plate was rotated by $90\degr$ after each subexposure.
All spectra have a resolving power of about 110,000 and cover the spectral
range 3780--6910\,\AA{}, with a small gap between 5259\,\AA{} and 5337\,\AA{}.
The reduction and calibration of these spectra was performed using the HARPS
data reduction software available on La~Silla. The normalization of the
spectra to the continuum level is described in detail in
\citet{Hubrig2013}.
A summary of all HARPS\-pol observations is given in Table~\ref{tab:log}. 
The columns list the heliocentric Julian date (HJD) for the middle of the 
subexposures, the exposure time, and the signal-to-noise ratio (S/N) of the
spectra. 

Due to the rather short pulsation period of HD\,21190 of the order of 3.6\,h
discovered by
\citet{Koen2001},
the shape of the line profiles in the high-resolution spectropolarimetric
observations is changing significantly over the pulsational cycle. Recently,
\citet{hd96446}
showed that such high-resolution observations of pulsating variable stars
frequently fail to deliver credible measurement results, if the whole sequence
of subexposures at different retarder waveplate angles has a duration
comparable to the timescale of the pulsation variability. To minimize the
impact of pulsations on the magnetic field measurements, we asked for
observing time with FORS\,2 mounted on the 8\,m Antu telescope of the Very
Large Telescope. FORS\,2 is a multi-mode instrument equipped with polarisation
analysing optics comprising super-achromatic half-wave and quarter-wave phase
retarder plates, and a Wollaston prism with a beam divergence of 22$\arcsec$
in standard resolution mode. The exposure time with such an instrument is only
of the order of a couple of minutes and no impact of pulsation on the field
measurements is expected.

\begin{table}
\centering
\caption{
  Logbook of the FORS\,2 observations of HD\,21190 (left) and
  CPD\,$-$83\degr\,64B (right). The columns list the modified Julian date of
  mid-exposure and the achieved S/N in the Stokes $I$ spectra around
  5400\,\AA{}.
}
\label{tab:fors2}
\begin{tabular}{crp{2mm}cr}
\hline
\multicolumn{2}{c}{HD\,21190} &
&
\multicolumn{2}{c}{CPD\,$-$83\degr\,64B} \\
\multicolumn{1}{c}{MJD} &
\multicolumn{1}{c}{S/N} &
&
\multicolumn{1}{c}{MJD} &
\multicolumn{1}{c}{S/N} \\
\hline
58003.2985 & 1665 & & 58003.2660 & 1205  \\
58031.1233 & 3134 & & 58031.1826 & 1200 \\
58033.0411 & 3107 & & 58035.2615 & 1355   \\
58035.2202 & 4117 & & 58049.3210 &  955  \\
58044.2433 & 3400 & & 58080.0461 &  930   \\
58116.0827 & 3500 & & 58116.1296 &  1200  \\
58150.0854 & 2570 \\
58162.0494 & 2520 \\
58207.0463 & 2540 \\
\hline
\end{tabular}
\end{table}

The acquired nine observations of HD\,21190 and six observations of
CPD\,$-$83\degr\,64B  summarized in Table~\ref{tab:fors2} were spread between
2017 September and 2018 February. The columns list the modified Julian dates
(MJD) for the middle of the exposure and the S/N of the spectra. We used 
the GRISM 600B and the narrowest available slit width of 0$\farcs$4 to obtain
a spectral resolving power of $R\approx2000$. The observed spectral range from
3250 to 6215\,\AA{} includes all Balmer lines, apart from H$\alpha$, and
numerous helium lines. Further, in our observations we used a non-standard
readout mode with low gain (200kHz,1$\times$1,low), which provides a broader
dynamic range, hence allowing us to reach a higher signal-to-noise ratio (S/N)
in the individual spectra. The position angle of the retarder waveplate was
changed from $+45^{\circ}$ to $-45^{\circ}$ and vice versa every second
exposure, i.e. we have executed the sequence $+45^{\circ}$$-45^{\circ}$,
$-45^{\circ}$$+45^{\circ}$, $+45^{\circ}$$-45^{\circ}$, etc.\ up to ten times for
HD\,21190 and four times for CPD\,$-$83\degr\,64B. Using this sequence of the
retarder waveplate ensures an optimum removal of instrumental polarization.
The exposure time for observations of each pair of subexposures
$+45^{\circ}$$-45^{\circ}$ or $-45^{\circ}$$+45^{\circ}$, including overheads,
accounted for about 4\,min for HD\,21190 and about 12\,min for
CPD\,$-$83\degr\,64B.


\section{Magnetic field analysis using high-resolution HARPS spectropolarimetry}
\label{sec:mfield}

\begin{figure*}
 \centering
        \includegraphics[width=0.8\textwidth]{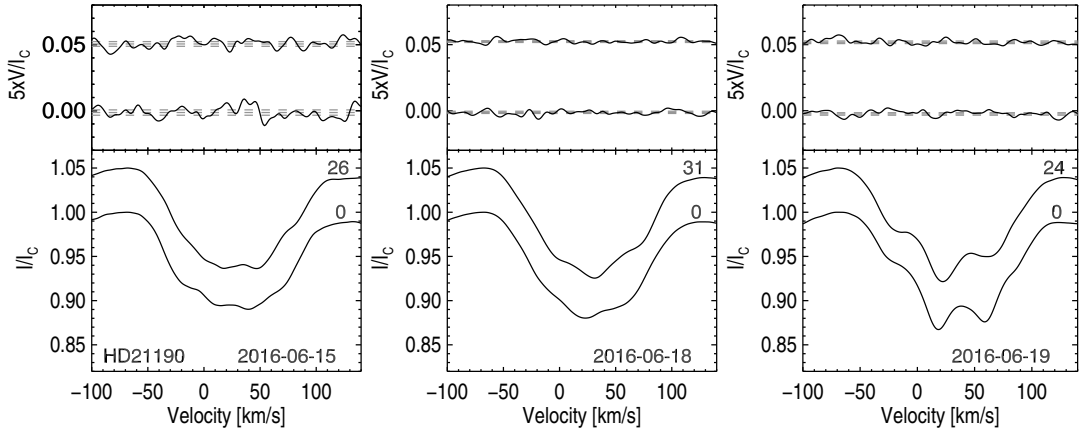}
       \caption{
         LSD Stokes~$V$ (top) and $I$ (bottom) profiles calculated
         for the individual subexposures obtained in the HARPS\-pol
         observations of HD\,21190 of three different nights. The numbers next
         to the profiles indicate the time lapse between the first and
         subsequent subexposures in minutes.
       }
   \label{fig:priprof}
\end{figure*}

To study the presence of a magnetic field in HD\,21190 and 
CPD\,$-$83$^{\circ}$\,64B we employed the Least-Squares Deconvolution (LSD)
technique
\citep{Donati1997}.
This technique combines line profiles (using the assumption that line
formation is similar in all lines) centred at the position of the individual
lines given in the line mask and scaled according to the line strength and
sensitivity to a magnetic field (i.e.\ to a Land{\'e} factor). The resulting
average profiles (Stokes $I$, Stokes $V$, and $N$) obtained by combining several
lines, yield an increase in S/N as the square root of the number of lines used 
and therefore in sensitivity to polarization signatures. The diagnostic $N$
profiles are usually used to identify spurious polarization signatures. They
are calculated by combining the subexposures in such a way that the
polarization cancels out.

Line masks for each component were constructed using the Vienna Atomic Line
Database
\citep[VALD; e.g.][]{Kupka2011,VALD3}
based on the stellar parameters of both targets. Only the lines that are
present in the stellar spectra are used. Further, obvious line blends and
lines in telluric regions were excluded from the line list. The mean 
longitudinal magnetic field is evaluated by computing the first-order moment 
of the Stokes~$V$ profile according to
\citet[][]{Mathys1989}:

\begin{equation}
\left<B_{\mathrm z}\right> = -2.14 \times 10^{11}\frac{\int \upsilon V
  (\upsilon){\mathrm d}\upsilon }{\lambda_{0}g_{0}c\int
  [I_{c}-I(\upsilon )]{\mathrm d}\upsilon},
\end{equation}

\noindent
where $\upsilon$ is the Doppler velocity in \kms, and $\lambda_{0}$ and
$g_{0}$ are the mean values for the wavelength (in nm) and the Land\'e factor
obtained from all lines used to compute the LSD profile, respectively. We note
that this equation is valid only in the velocity domain.

For HD\,21190 we constructed a line mask containing 156 metallic lines with a
mean Land\'e factor $\bar{g}_{\rm eff} =1.32$ based on the stellar parameters
$T_\mathrm{eff}=6900\pm100$~K and $\log g=3.6\pm0.2$ obtained by
\citet{niemczura}.
Since the subexposure times of 23--32\,min of the HARPS\-pol observations of
this $\delta$~Scuti variable are rather long, constituting a significant
fraction of the pulsation cycle, the observed line profiles exhibit very strong
pulsational variability, leading to pulsational wavelength shifts that affect
the magnetic field measurements. The impact of the pulsations is clearly visible
in Fig.~\ref{fig:priprof}, where we present the LSD Stokes~$V$ and $I$ profiles
calculated for individual subexposures. 

To exclude the pulsational impact on the magnetic field measurements, we
decided to carry out the measurements of the line shifts between the spectra
$(I+V)_{\rm 0}$ and $(I-V)_{\rm 0}$ in the first subexposure and the spectra
$(I+V)_{\rm 90}$ and $(I-V)_{\rm 90}$ in the second subexposure separately 
\citep[e.g.][]{Hubrig2011, hd96446}.
No definite magnetic field detection with a false alarm probability (FAP)
$<$\,$10^{-5}$ was achieved in the HARPS\-pol observations of HD\,21190. Only
for the first subexposure obtained during the first observing night we obtain
$\left< B_{\rm z}\right>=87\pm33$~G with a false alarm probability (FAP) of
$5.8\times10^{-4}$, indicating a marginal detection. We classify the magnetic
field measurements making use of the FAP
\citep{Donati-fap},
considering a profile with FAP\,$<$\,$10^{-5}$ as a definite detection,
$10^{-5}$\,$\le$\,FAP\,$<10^{-3}$ as a marginal detection, and
FAP\,$\ge$\,$10^{-3}$ as a non-detection.

\begin{figure}
 \centering
        \includegraphics[width=.7\columnwidth]{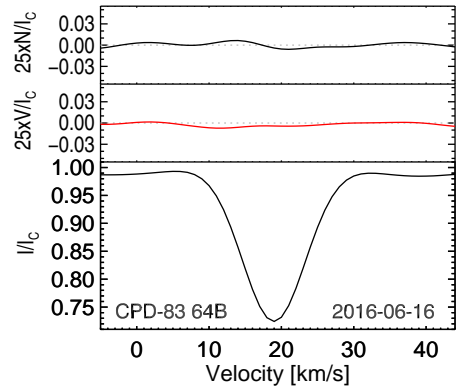}
        \caption{LSD Stokes $I$, $V$, and $N$ profiles of
          CPD\,$-$83\degr\,64B combining all four subexposures.
                }
   \label{fig:secfin}
\end{figure}

\begin{figure}
 \centering
        \includegraphics[width=0.85\columnwidth]{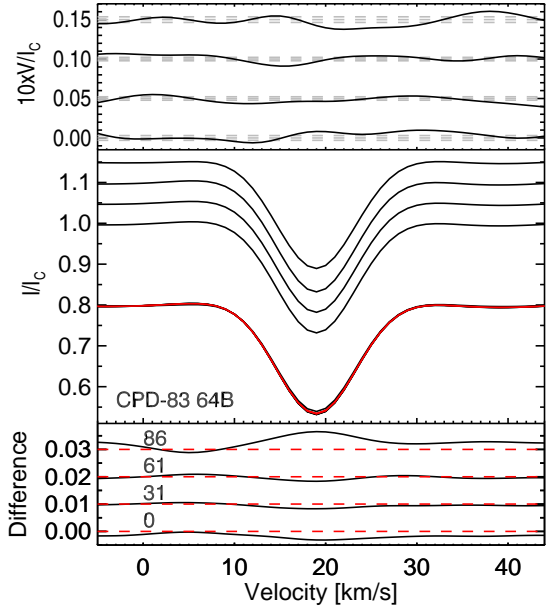}
        \caption{
          \emph{Top:} LSD Stokes $V$ profiles of CPD\,$-$83\degr\,64B
          calculated for each individual subexposure. The subsequent profiles
          are shifted vertically for better visibility.
          \emph{Middle:} Individual LSD Stokes~$I$ profiles 
          compared with the average Stokes~$I$ profile (red line). The
          profiles are also shifted vertically for better visibility.
          \emph{Bottom:} Differences between the Stokes~$I$ profiles computed
          for the individual subexposures and the average Stokes~$I$ profile.
          The dashed lines were added to guide the eye. The time difference in
          minutes to the first subexposure is given next to the profiles.
                }
   \label{fig:secpuls}
\end{figure}

The chemical abundance analysis of CPD\,$-$83$\bmath{\degr}$64B by
\citet{niemczura}
revealed that it is a G-type star with $T_\mathrm{eff}=5850\pm50$~K and
$\log g=4.3\pm0.1$. The LSD Stokes~$I$ and $V$ profile, together with the
diagnostic null profile $N$, obtained using 730 metallic lines with a mean
Land\'e factor $\bar{g}_{\rm eff} =1.27$ are presented in Fig.~\ref{fig:secfin}.
However, as Fig.~\ref{fig:secfin} shows, the LSD null profile $N$ does not
appear flat, indicating a possible presence of variability also in this star.
Similar features were also observed in the null profiles of other G-type stars
\citep[see e.g.][]{Marsden}.
Therefore, to minimize the impact of pulsations on the magnetic field
measurements, we analysed the presence of a magnetic field on each individual
subexposure in a similar way as for HD\,21190. In Fig.~\ref{fig:secpuls} we
present the LSD Stokes~$I$ and $V$ profiles of CPD\,$-$83\degr\,64B for each
individual subexposure and the comparison of the individual LSD Stokes~$I$
profiles with the average LSD Stokes $I$ profile. The distinct line profile
variability, probably caused by the presence of pulsations, is clearly visible
in the bottom panel displaying differences between the Stokes~$I$ profiles
computed for the individual subexposures and the average Stokes~$I$ profile.
As we show later in Sect.~\ref{sec:per}, the rotation period of
CPD\,$-$83\degr\,64B is too long for variability to be caused by spots.

Only marginal field detections were achieved for the individual subexposures
with field strengths varying from $-103$\,G to 218\,G, the best FAP value
attained was $4\times10^{-5}$. To summarize, our high-resolution HARPS\-pol
observations with long exposure times are certainly affected by the presence
of pulsations and do not allow us to make any definite conclusion on the
presence or absence of a longitudinal magnetic field in either star.

\section{Magnetic field analysis using low-resolution FORS\,2 spectropolarimetry}
\label{sec:fors}

A description of the assessment of the presence of a longitudinal magnetic
field using FORS\,1/2 spectropolarimetric observations was presented in our
previous work
\citep[e.g.][and references therein]{Hubrig2004a, Hubrig2004b}.
Rectification of the $V/I$ spectra was performed in the way described by
\citet{HubrigFORS2}.
Null profiles $N$ are calculated as pairwise differences from all available
$V$ profiles so that the real polarisation signal should cancel out. From
these, 3$\sigma$-outliers are identified and used to clip the $V$ profiles.
This removes spurious signals, which mostly come from cosmic rays, and also
reduces the noise. A full description of the updated data reduction and
analysis will be presented in a separate paper
\citep[Sch\"oller et al., in preparation, see also][]{HubrigFORS2}.

The mean longitudinal magnetic field, $\left< B_{\rm z}\right>$, is measured on
the rectified and clipped spectra based on the relation following the method
suggested by
\citet{angelland}:

\begin{eqnarray}
\frac{V}{I} = -\frac{g_{\rm eff}\, e \,\lambda^2}{4\pi\,m_{\rm e}\,c^2}\,
\frac{1}{I}\,\frac{{\rm d}I}{{\rm d}\lambda} \left<B_{\rm z}\right>\, ,
\label{eqn:vi}
\end{eqnarray}

\noindent
where $V$ is the Stokes parameter that measures the circular polarization,
$I$ is the intensity in the unpolarized spectrum, $g_{\rm eff}$ is the effective
Land\'e factor, $e$ is the electron charge, $\lambda$ is the wavelength,
$m_{\rm e}$ is the electron mass, $c$ is the speed of light,
${{\rm d}I/{\rm d}\lambda}$ is the wavelength derivative of Stokes~$I$, and
$\left<B_{\rm z}\right>$ is the mean longitudinal (line-of-sight) magnetic
field.

\begin{table}
\centering
\caption{Longitudinal magnetic field values obtained for HD\,21190 using
  FORS\,2 observations. In the first column we show the modified Julian dates
  of mid-exposures, followed by the mean longitudinal magnetic field using the
  Monte Carlo bootstrapping test, for all lines, and its significance.
  In the two last columns, we present the results of our measurements using
  only the hydrogen lines and when using the null spectra.
  All quoted errors are 1$\sigma$ uncertainties.
}
\label{tab:priMFFORS}
\begin{tabular}{lr@{$\pm$}lcr@{$\pm$}lr@{$\pm$}l}
\hline
\multicolumn{1}{c}{MJD} &
\multicolumn{2}{c}{$\left<B_{\rm z}\right>_{\rm all}$} &
\multicolumn{1}{c}{Signif.} &
\multicolumn{2}{c}{$\left<B_{\rm z}\right>_{\rm hyd}$} &
\multicolumn{2}{c}{$\left<B_{\rm z}\right>_{\rm N}$} \\
&
\multicolumn{2}{c}{(G)} &
\multicolumn{1}{c}{$\sigma$} &
\multicolumn{2}{c}{(G)} &
\multicolumn{2}{c}{(G)}
\\
\hline
54343.2791$^1$ & \multicolumn{2}{c}{-} & -   & 47     & 13 &  \multicolumn{2}{c}{-} \\
57462.0051$^2$ & $-$254 & 60 & 4.3 & $-$237 & 75 &   54 & 62 \\
58003.2985     & 246    & 61 & 4.0 & 174    & 91 &   19 & 57 \\
58031.1233     & 129    & 44 & 2.9 & 96     & 68 &$-$39 & 35 \\
58033.0411     & 257    & 64 & 4.0 & 200    & 97 &   26 & 44 \\
58035.2202     & 173    & 35 & 4.9 & 148    & 54 &    5 & 24 \\
58044.2433     & 230    & 38 & 6.1 & 132    & 61 &   43 & 30 \\
58116.0827     & 137    & 35 & 3.9 & 159    & 57 &$-$21 & 27 \\
58150.0854     & 46     & 39 & 1.2 & 45     & 62 & 29   & 41 \\
58162.0494     & $-$15  & 27 & 0.6 & $-$23  & 66 &$-$3  & 28 \\
58207.0463     & 160    & 62 & 2.6 & 211    & 96 &$-$9  & 60 \\
\hline
\end{tabular}
\begin{minipage}{\textwidth}
{\bf Note:}\\
$^1$ {The measurement at MJD 54343.2791 was previously reported \\ by \citet{Kurtz2008}}.\\
$^2$ {The measurement at MJD 57462.0051 was previously reported \\ by \citet{2016HubSch}}.\\
\end{minipage}
\end{table}

\begin{table}
\centering
\caption{Longitudinal magnetic field values obtained for CPD\,$-$83\degr\,64B
  using FORS\,2 observations. In the first column we show the modified Julian
  dates of mid-exposure, followed by the mean longitudinal magnetic field
  using the Monte Carlo bootstrapping test, for all lines, and its
  significance.
  In the two last columns, we present the results of our measurements using
  only the hydrogen lines and when using the null spectra.
  All quoted errors are 1$\sigma$ uncertainties. 
}
\label{tab:secMFFORS}
\begin{tabular}{lr@{$\pm$}lcr@{$\pm$}lr@{$\pm$}l}
\hline
\multicolumn{1}{c}{MJD} &
\multicolumn{2}{c}{$\left<B_{\rm z}\right>_{\rm all}$} &
\multicolumn{1}{c}{Signif.} &
\multicolumn{2}{c}{$\left<B_{\rm z}\right>_{\rm hyd}$} &
\multicolumn{2}{c}{$\left<B_{\rm z}\right>_{\rm N}$} \\
&
\multicolumn{2}{c}{(G)} &
\multicolumn{1}{c}{$\sigma$} &
\multicolumn{2}{c}{(G)} &
\multicolumn{2}{c}{(G)} \\
\hline
58003.2660 & 375    & 99  & 3.8 & 315    & 177 &  55 & 85 \\
58031.1826 & 186    & 114 & 1.6 & 270    & 176 &$-$7 &112 \\
58035.2615 & 509    & 104 & 4.9 & 435    & 152 &$-$11& 83 \\
58049.3210 & 157    & 169 & 1.2 & 311    & 261 &$-$19&155 \\
58080.0461 & $-$225 & 64  & 3.5 & $-$253 & 111 &$-$17&74  \\
58116.1296 & 322    & 108 & 3.0 & 185    & 187 &$-$47&92  \\
\hline
\end{tabular}
\end{table}

\begin{figure}
 \centering
 \includegraphics[width=\columnwidth]{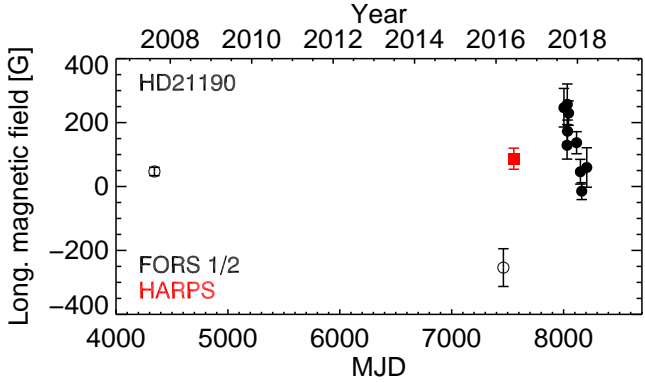}
 \caption{Distribution of the mean longitudinal magnetic field values of
   HD\,21190 as a function of MJD between 2007 and 2018. The black open
   (already published) and filled (new) circles represent FORS\,2
   observations, whereas the red filled square is used for the HARPS\-pol
   observation on 2016 June 15.}
 \label{fig:priB}
\end{figure}

\begin{figure}
 \centering
        \includegraphics[width=\columnwidth]{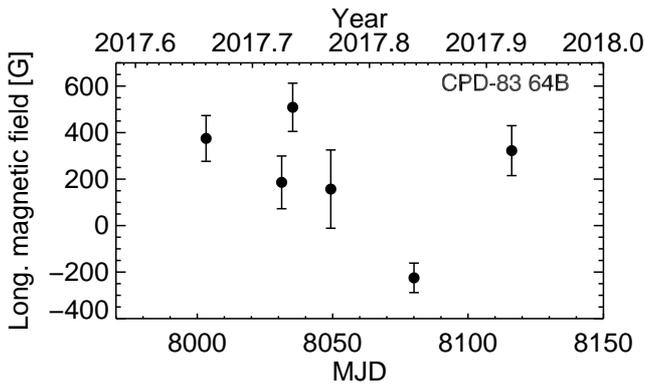}
        \caption{Distribution of the mean longitudinal magnetic field values
          of CPD\,$-$83\degr\,64B as a function of MJD in 2017 and 2018. 
                }
   \label{fig:secB}
\end{figure}

The longitudinal magnetic field was measured in two ways: using the entire
spectrum including all available lines, or using exclusively hydrogen lines.
Furthermore, we have carried out Monte Carlo bootstrapping tests. These are
most often applied with the purpose of deriving robust estimates of standard
errors
\citep[e.g.][]{Steffen}. 
The measurement uncertainties obtained before and after the Monte Carlo
bootstrapping tests were found to be in close agreement, indicating the
absence of reduction flaws. The results of our magnetic field measurements,
those for the entire spectrum or only for the hydrogen lines are presented for
HD\,21190 and CPD\,$-$83\degr\,64B in Tables~\ref{tab:priMFFORS} and
\ref{tab:secMFFORS}, respectively. Distributions of the mean longitudinal
magnetic field values as a function of MJD obtained for both stars are
presented in Figs.~\ref{fig:priB} and ~\ref{fig:secB}.

For HD\,21190, the values for the longitudinal magnetic field
$\left< B_{\rm z}\right>_{\rm all}$ vary roughly between $-250$\,G and 250\,G,
with the measurement $\left< B_{\rm z}\right>_{\rm all}=230\pm38$\,G at a
significance level of 6.1$\sigma$. The magnetic field appears to be stronger in
CPD\,$-$83\degr\,64B although the uncertainties of the field determination in
this component are higher due to its faintness with $m_{\rm v}=10.8$. The
highest value for the longitudinal magnetic field,
$\left< B_{\rm z}\right>_{\rm all}=509\pm104$\,G, is measured at a significance
level of 4.9$\sigma$. 

\section{Period analysis}
\label{sec:per}

Based on the new distance estimates presented in Table~\ref{tab:table_gaiadr2}
and the apparent magnitudes of the stars
\citep[the ASCC-2.5 catalogue;][$V=7.604$ for HD\,21190 and $V=10.815$ for CPD\,$-$83\degr\,64B]{kharchenko01},
we obtain for both targets the absolute visual magnitudes $M_{V}=1.10\pm0.05$
and $M_{V}=3.85\pm0.03$, respectively. After applying the corresponding
bolometric corrections of 0.028 and $-$0.067
\citep{flower},
we obtain the luminosity $\log (L/L_{\sun})=1.44\pm0.01$ for HD\,21190 and
the luminosity $\log (L/L_{\sun})=0.38\pm0.01$ for CPD\,$-$83\degr\,64B.
Using the Stefan-Boltzmann law with the effective temperature values derived
from high-resolution spectra by
\citet{niemczura},
we derive stellar radii of $3.69\pm0.07$\,$R_{\sun}$ and
$1.51\pm0.03$\,$R_{\sun}$. The obtained radii are identical within the
uncertainties with the radii estimated using the \emph{Gaia} DR2 data
presented in Table~\ref{tab:table_gaiadr2}. With these radii and the
$v \sin i$ values of $74\pm3$\,\kms{} and $3.1\pm0.7$\,\kms{} from
\citet{niemczura},
we obtain upper limits for the rotation period:
$P_{\rm rot} \le 2.5 \pm 0.1$\,d for HD\,21190 and
$P_{\rm rot} \le 25 \pm 4$\,d for CPD\,$-$83\degr\,64B.

 \begin{figure*}
 \centering
 \includegraphics[width=\columnwidth]{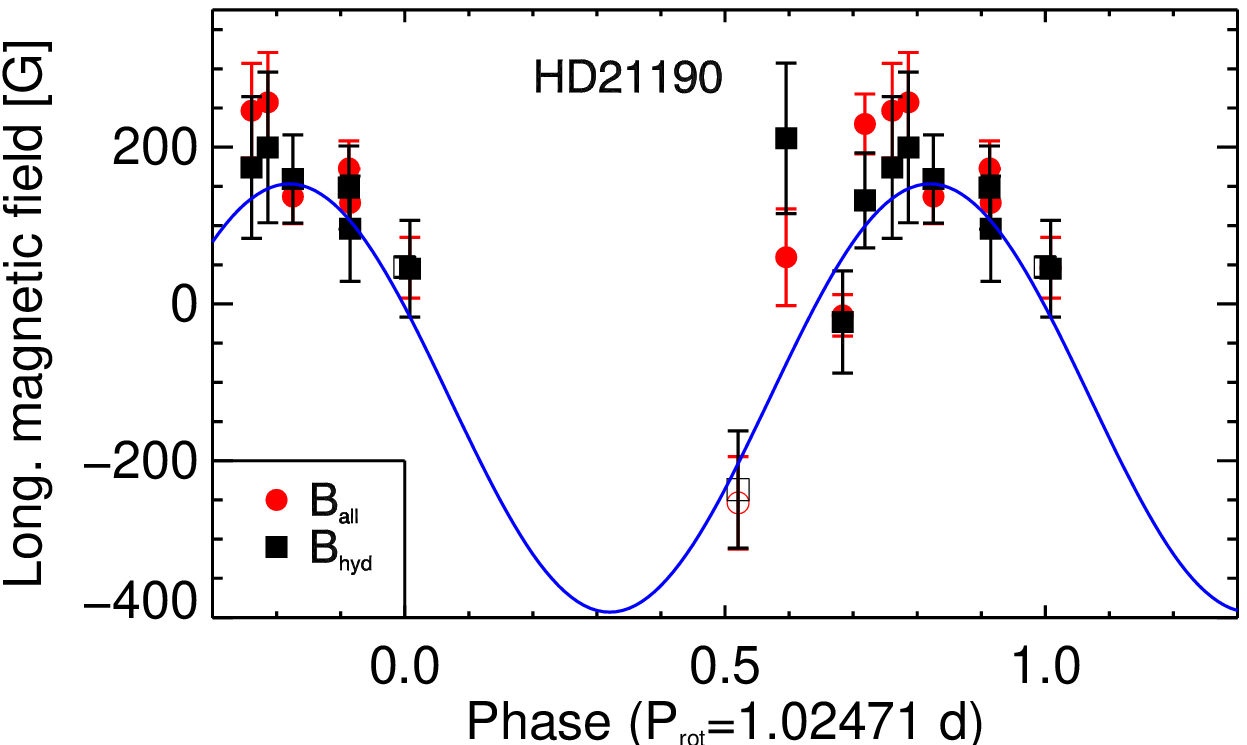}
\includegraphics[width=\columnwidth]{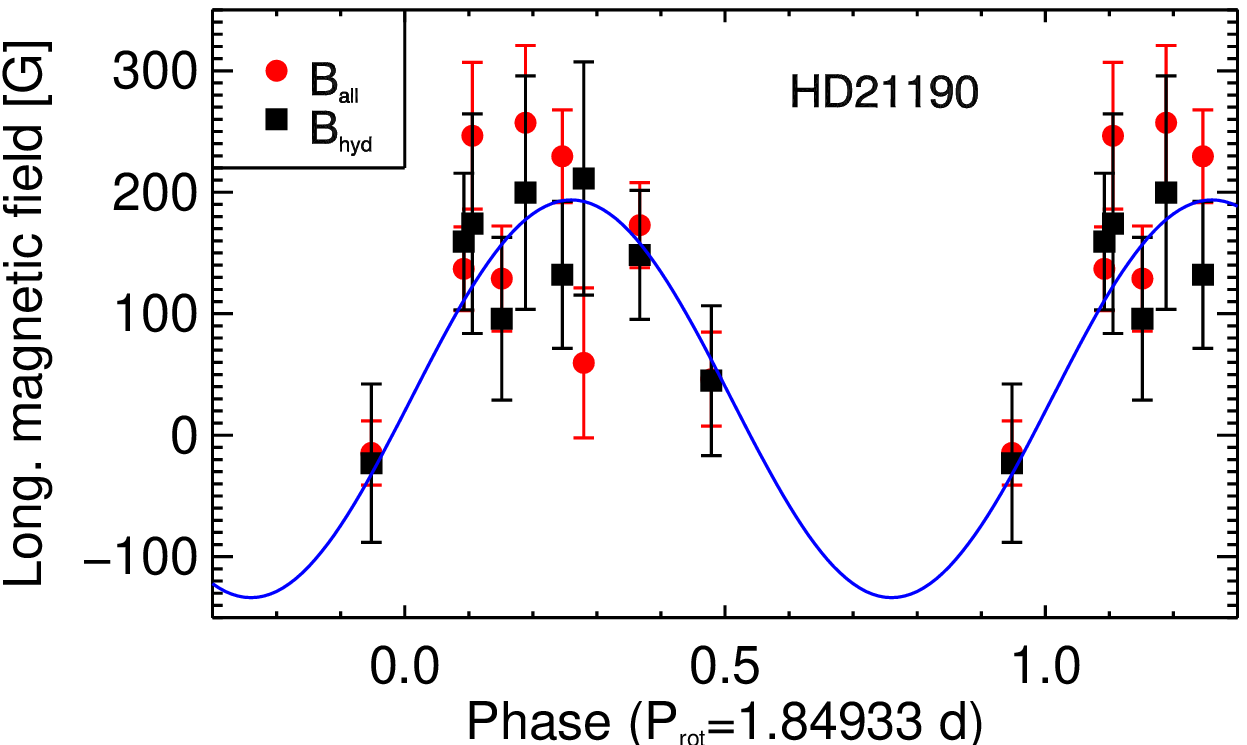}
\caption{{\it Left panel:} Distribution of the mean longitudinal magnetic
  field values of HD\,21190 phased with the rotation period
  $P_{\rm rot} = 1.02471\pm0.00045$\,d. The black filled squares represent
  FORS\,2 measurements using the hydrogen lines and the red filled circles the
  measurements using the entire spectrum. Previously published measurements
  are presented by open symbols.
  {\it Right panel:} Distribution of the mean longitudinal magnetic field
  values of HD\,21190 phased with the rotation period
  $P_{\rm rot} = 1.84933\pm0.00029$\,d.}
 \label{fig:mf}
\end{figure*}

 Magnetic Ap and Bp stars usually exhibit photometric, spectral and magnetic
 variability with the rotation period. Unfortunately, in our study, the number
 of available magnetic field measurements is only eleven (including two
 previously published measurements) for the primary component and only six for
 the secondary component. We nevertheless tried to detect the most probable
 rotation period for HD\,21190 by calculating the frequency spectrum. Our
 analysis was performed using a non-linear least squares fit to the multiple
 harmonics utilizing the Levenberg-Marquardt method
 \citep{press}.
 For each trial frequency we performed a statistical F-test of the null
 hypothesis for the absence of periodicity
 \citep{seber}.
 The resulting F-statistics can be thought of as the total sum including
 covariances of the ratio of harmonic amplitudes to their standard deviations,
 i.e.\ a signal-to-noise ratio.

 The highest peak in the periodogram calculated for the magnetic field
   measurements including the previously published values corresponds to a
   period of $1.02471\pm0.00045$\,d with an FAP of $2.89\times10^{-9}$. The
   distribution of the mean longitudinal magnetic field values phased with
   this period is presented in Fig.~\ref{fig:mf} in the left panel. Using the
   above estimated stellar radius $R= 3.69\pm 0.07\,R_{\sun}$,
   $v \sin i = 74\pm 3$\,km\,s$^{-1}$, and the rotation period
   $P_{\rm rot} = 1.02471\pm0.00045$\,d, we obtain
   $v_{\rm eq}=182\pm4$\,km\,s$^{-1}$ and an inclination angle of the stellar
   rotation axis to the line of sight $i=24\pm2^{\circ}$.

In the case when only the new measurements of the magnetic field are used, the 
highest peak in the periodogram corresponds to a period of
$1.84933\pm0.00043$\,d with an FAP of $2.01\times10^{-6}$. The distribution of
the most recent mean longitudinal magnetic field values phased with this
period is presented in Fig.~\ref{fig:mf} in the right panel. This period
leads to $v_{\rm eq}=101\pm2$\,km\,s$^{-1}$ and an inclination angle of the
stellar rotation axis to the line of sight $i=47\pm1^{\circ}$.

We note that due to the rather low number of magnetic field measurements, our
results on the period search are only tentative and should be verified by
additional spectropolarimetric observations.

\section{Discussion}
\label{sec:disc}

The recent atmospheric chemical analysis of
\citet{niemczura}
based on HARPS\-pol observations confirmed the presence of chemical
peculiarities in HD\,21190. According to the authors, HD\,21190 is currently
the only known case of an Ap star with a detected magnetic field that shows
$\delta$\,Scuti pulsations. CPD\,$-$83\degr\,64B appeared mostly solar with
modest enhancements of heavy and rare-earth elements.

Although previous studies showed that HD\,21190 exhibits $\delta$\,Scuti
pulsations, the possible presence of pulsations in CPD\,$-$83$^{\circ}$\,64B
was discovered in the current work exclusively through spectroscopic
observations. Pulsations in solar-like stars are mainly acoustic modes excited
by turbulent convection like in the Sun, and are predicted for all
main-sequence and post-main-sequence stars cool enough to harbour an outer
convective envelope
\cite[e.g.][]{dimauro}.
Future spectroscopic and photometric monitoring will be worthwhile to
characterize this short-time variability in CPD\,$-$83$^{\circ}$\,64B.

Whereas the previous astrometric analysis
\citep{niemczura}
using {\it Gaia} DR1 data
\citep{gaia16}
showed that the $\delta$~Scuti variable HD\,21190 forms a physical binary
system with CPD\,$-$83\degr\,64B at an angular distance of about 19\arcsec,
the more recent {\it Gaia} DR2 results clearly indicate that the two stars
are not physically associated.

The initial idea to search for magnetic fields in HD\,21190 and
CPD\,$-$83\degr\,64B originated under the aspect that both targets are
physical components of a wide binary.
A detection of magnetic fields in components of a wide binary could possibly
be related to the most important issue related to the nature of magnetic
chemically peculiar stars, namely the origin of their magnetic fields, which
is still not properly understood. A scenario for the origin of the magnetic
fields of Ap stars, in which these stars result from the merging of two lower
mass protostars, was suggested by
\citet{Ferrario}.
In this scenario, the binaries that are observed now were triple systems earlier
in their history. If the hypothesis that the magnetic field originates in
dynamos driven by violent mixing of stellar plasma when two stars merge is
valid, one would expect that only very few magnetic massive stars are found in
close binaries. Indeed, close binary or multiple systems with the particular
combination of a magnetic and a non-magnetic star or two magnetic stars are
very rare. Among the studied binary systems with A and late B-type primaries,
only two systems, HD\,98088 and HD\,161701, are known to have
a magnetic Ap star as a component
\citep{Babcock, Hubrig2014}.
Should a wide third component be essential to lead to mergers and thus to Ap
stars with a magnetic field, this wide component should still be physically
bound to the magnetic Ap star and be detectable as a wider binary component
\citep{Mazeh}.
Unfortunately, the new astrometric analysis of HD\,21190 and
CPD\,$-$83\degr\,64B showed that these two objects are not such a wide pair.

\section*{Acknowledgements}

We are grateful for very helpful comments from an anonymous referee.
Based on observations made with ESO Telescopes at the La Silla Paranal
Observatory under programme IDs 079.D-0241 (PI:Briquet), 191.D-0255
(PI: Morel), 097.C-0277 (PI: Hubrig), and 0100.D-0137 (PI: J\"arvinen).
This work has made use of the VALD database, operated at Uppsala
University, the Institute of Astronomy RAS in Moscow, and the University of
Vienna.
EN acknowledges the Polish National Science Center grants no.\ 
2014/13/B/ST9/00902.








\bsp	
\label{lastpage}
\end{document}